\documentclass[preprint,aps,showpacs]{revtex4}

\usepackage{amsmath,amssymb}
\usepackage{graphicx}

\newcommand{\<}{\langle}
\renewcommand{\>}{\rangle}
\newcommand{\be}{\begin{equation}}
\newcommand{\ee}{\end{equation}}
\newcommand{\bea}{\begin{eqnarray}}
\newcommand{\eea}{\end{eqnarray}}

\renewcommand{\a}{\alpha}

\newcommand{\uh}{\widehat{u}}
\newcommand{\Ch}{\widehat{C}}
\newcommand{\Fh}{\widehat{F}}
\newcommand{\Ph}{\widehat{P}}
\newcommand{\Qh}{\widehat{Q}}
\newcommand{\Q}{Q}
\newcommand{\deh}{\widehat{\delta}}
\newcommand{\de}{\delta}
\newcommand{\rh}{\widehat{\rho}}
\renewcommand{\r}{\rho}
\newcommand{\eh}{\widehat{\epsilon}}
\newcommand{\e}{\epsilon}
\newcommand{\rr}{\widehat{r}}
\newcommand{\Th}{\widehat{T}}
\renewcommand{\th}{\widehat{t}}
\newcommand{\us}{\underline{\sigma}}
\newcommand{\da}{\partial a}
\newcommand{\di}{\partial i}
\newcommand{\ex}{\text{e}}

\begin{document}

\title{Instability of one-step replica-symmetry-broken phase in
satisfiability problems}

\author{Andrea Montanari}

\affiliation{Laboratoire de Physique Th\'{e}orique de l'Ecole Normale
Sup\'{e}rieure\footnote {UMR 8549, Unit{\'e} Mixte de Recherche du
Centre National de la Recherche Scientifique et de l' Ecole Normale
Sup{\'e}rieure. }, 24, rue Lhomond, 75231 Paris CEDEX 05, France}

\author{Giorgio Parisi and Federico Ricci-Tersenghi}

\affiliation{Dipartimento di Fisica, INFM (UdR and SMC centre) and
INFN, Universit\`{a} di Roma ``La Sapienza'', Piazzale Aldo Moro 2,
I-00185 Roma, Italy}

\date{\today}

\begin{abstract}
We reconsider the one-step replica-symmetry-breaking (1RSB) solutions
of two random combinatorial problems: $k$-XORSAT and $k$-SAT.  We
present a general method for establishing the stability of these
solutions with respect to further steps of replica-symmetry
breaking. Our approach extends the ideas of Ref.~\cite{NostroInst} to
more general combinatorial problems.

It turns out that 1RSB is {\it always} unstable at sufficiently small
clauses density $\alpha$ or high energy. In particular, the recent
1RSB solution to 3-SAT is unstable at zero energy for
$\alpha<\alpha_{\rm m}$, with $\alpha_{\rm m}\approx 4.153$. On the
other hand, the SAT-UNSAT phase transition seems to be correctly
described within 1RSB.
\end{abstract}


\maketitle

\section{Introduction}
\label{Introduction}

It is well known that there are two possible structures for the
low-temperature phase of mean-field spin glasses \cite{SpinGlass}.
The first scenario is described, within replica theory, by a 1RSB
ansatz. It corresponds to the existence of an exponential number of
pure states which are, roughly speaking, uncorrelated. In the second
scenario, a large number~\footnote{An estimate of their number is
still matter of debate.  See Refs. \cite{Bray,Cavagna,Crisanti} for
some recent contributions.}  of pure states is organized in an
ultrametric tree. The tree describes the probabilistic dependencies
among the free energies and the distances of different pure states.
This probabilistic structure corresponds, in replica jargon, to a full
replica symmetry breaking (FRSB) ansatz.

In the last 20 years many combinatorial problems have been
successfully analyzed using the well known mapping onto disordered
statistical physics models \cite{SpinGlass,CSReview}.  The same two
scenarios are expected to be present within this domain.  However,
because of the rich structure of many combinatorial problems, their
analysis has been so-far limited to 1RSB calculations.  It is
therefore of the utmost importance to analyze the consistency of the
1RSB solutions. An important check consists in looking at a
neighborhood of the 1RSB subspace (in the larger FRSB space) and
verify the ``local stability'' of the 1RSB solution.

Such a computation has recently acquired a further reason of
interest. As shown in Ref.~\cite{NostroInst}, even in situations in
which the equilibrium behavior is correctly treated within a 1RSB
ansatz, non-equilibrium properties generically require a FRSB
description. In fact, it turns out that high-lying metastable states
are unstable towards FRSB. In a combinatorial optimization context
``equilibrium properties'' are related to the cost of the optimal
solution. ``Metastable states'' are, possibly, related to the dynamics
of local search algorithms \cite{PTAC}.

\begin{figure}
\begin{center}
\includegraphics[width=0.65\textwidth]{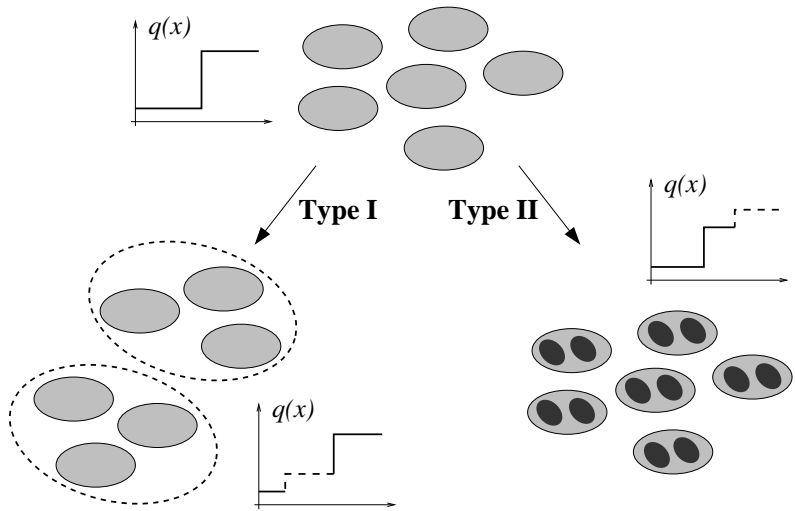}
\end{center}
\caption{A pictorial view of the two types of instabilities of the
1RSB solution. Blobs represent pure states or clusters of states.}
\label{InstabilityTypesFig}
\end{figure}
There are two possible instabilities of the 1RSB calculation
\cite{NostroInst}. If we think of the 1RSB solution as describing the
decomposition of the Gibbs measure in many, well-separated pure
states, the following scenarios are possible: (I) The states organize
themselves into clusters, forming an ultrametric FRSB structure; (II)
Each state splits into many sub-states forming an FRSB structure.  In
Fig.~\ref{InstabilityTypesFig} we present a pictorial interpretation
of these two instabilities.  Type-I instability usually occurs at low
energy, and often below the ground-state energy. In this case it is
irrelevant from a statistical physics point of view.  The calculation
of Ref.~\cite{GiorgioInst} for 3-SAT was related to type-I
instability, and confirmed the irrelevance of this phenomenon.
Type-II instability affects always the high-lying metastable states
and, in some regions of the phase diagram, even the ground state.
Here we shall compute the threshold for this type of instability.

In this paper, we shall focus on satisfiability problems, and in
particular treat the special cases of $k$-XORSAT and $k$-SAT.  These
are problems in which a large number of boolean variables have to be
fixed in such a way to satisfy a set of constraints (clauses).
$k$-SAT lies at the very heart of theoretical computer science. It is,
in fact, one of the first problems which have been proved to be
NP-complete (the first being its irregular version: SAT) \cite{Garey}.
A large effort has been devoted to the study of the SAT-UNSAT
transition in random $k$-SAT \cite{3SAT,Variational}.  The 1RSB
solution of this model has been a quite recent achievement
\cite{MarcGiorgioRiccardo,MarcRiccardo}.  $k$-XORSAT is somehow
simpler than $k$-SAT (both from the analytic and the algorithmic point
of view), while sharing a similar phase diagram
\cite{Creignou,XorSat}.  Interestingly, its 1RSB solution has been
proved to be correct in the zero-energy limit
\cite{CoccoEtAlXorSat,MezardEtAlXorSat}.

Since we are mainly interested in combinatorial problems, we shall
focus on zero-temperature statistical mechanics. In the main part of
our paper we fix $T=0$ since the beginning of our calculations.  It is
however instructive to solve the problem (in 1RSB approximation) at
finite temperature and let $T\to 0$ afterwards.  It turns out that the
instability of the $T=0$ 1RSB solution is reflected in an unphysical
behavior of the finite temperature solution in the $T\to 0$
limit. This provides an useful check of our calculations.

The paper is organized as follows. In Sec.~\ref{GeneralSection} we
explain how the instability of the 1RSB phase can be derived from an
analysis of the two-step replica symmetry breaking (2RSB) saddle-point
equations. We describe our method for a general satisfiability model.
In Sec.~\ref{NumericalSection} we specialize to two prototypical
cases: random $k$-XORSAT and random $k$-SAT, and show the results of a
numerical evaluation of the stability condition. We discuss the
consequences of our findings.  In Sec.~\ref{NonIntegerSection} we
consider the finite-temperature 1RSB solution. We compare the behavior
of this solution in the $T\to 0$ limit and the stability thresholds
computed in the previous Section.  In Appendix~\ref{ExplicitFormulae}
we collect the explicit formulae for the stability of $k$-XORSAT and
$k$-SAT.  Finally in App.~\ref{ExpansionXORSAT} we expand around the
dynamical transition of $k$-XORSAT, in order to have an analytical
characterization of this transition.
%
%
\section{The general approach}
\label{GeneralSection}

In this Section we consider a general model over $N$ Ising spins
$\sigma_i=\pm 1$, $i\in\{1,\dots, N\}$. The Hamiltonian is the sum of
$M=\alpha N$ terms, each one being a $k$-spin interaction (we shall be
interested in the case $k\ge 3$). We use the indices
$a,b,c\dots\in\{1,\dots M\}$ for the interactions, and denote by
$\partial a = \{i^a_1,\dots,i^a_k\}$ the set of sites entering in the
interaction $a$. Conversely $\di$ will be the set of interactions in
which $i$ participates. With these conventions the Hamiltonian reads
\begin{eqnarray}
H(\us) = \sum_{a=1}^M E_a(\us_{\da})\, ,\label{Hamiltonian}
\end{eqnarray}
where we used the vector notations $\us = (\sigma_1,\dots,\sigma_N)$
and $\us_{\da} = (\sigma_{i^a_1},\dots,\sigma_{i^a_k})$.  The
functions $E_a(\cdot)$ may (eventually) depend upon some quenched
random variables which we will not note explicitly. Each interaction
(clause) $E_a(\cdot)$ can take two values: either $0$ (the clause is
satisfied) or $2$ (unsatisfied).
\begin{figure}
\begin{center}
\includegraphics[width=0.65\textwidth]{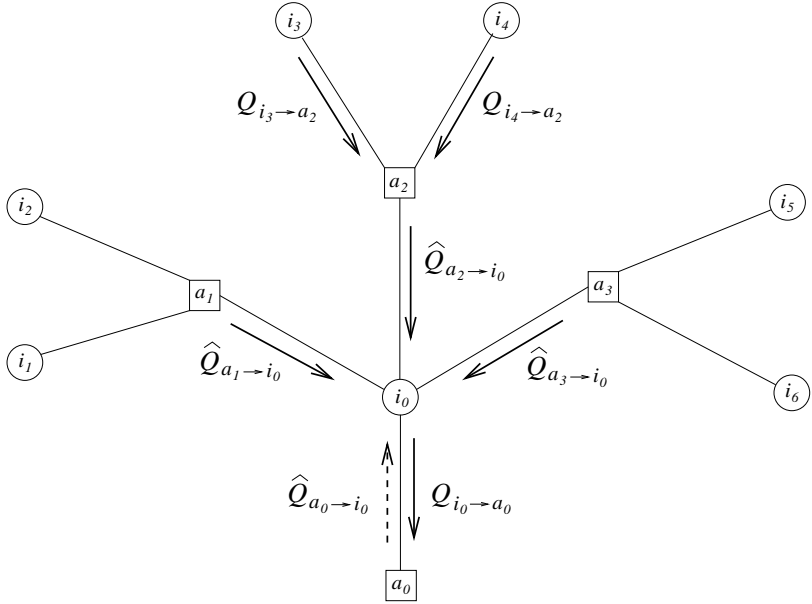}
\end{center}
\caption{A fragment of a factor graph. Squares represent clauses
(interactions) and circles represents variables (spins). Variable
nodes are connected by a link to the clauses they belong to. Cavity
fields (and their distributions) are associated with directed edges.}
\label{FactorFig}
\end{figure}

A nice graphical representation of such a model is obtained by drawing
a factor graph \cite{Factor}, cf.\ Fig.~\ref{FactorFig}.  This is a
bipartite graph with two type of nodes: variable nodes, and clause
nodes. An edge is drawn between the clause $a$ and the variable $i$ if
$i\in\da$ (or, equivalently, $a\in\di$).  A crucial property for mean
field theory to be exact is that the factor graph must not contain
``short'' loops.

Let ${\cal S}$ be the space of probability distribution $\rho \equiv
(\rho_+,\rho_0,\rho_-)$ over the set $\{+,0,-\}$.  Geometrically
${\cal S}$ is the two-dimensional simplex.  The zero-temperature 2RSB
order parameter for the model (\ref{Hamiltonian}) is given by a
distribution over ${\cal S}$ for each {\it directed} link of the
factor graph.  We shall denote such distributions by $\Q_{i\to a}[\r]$
(if the link is directed from a variable node to a clause node) or by
$\Qh_{a\to i}[\rh]$ (in the opposite case).  The 2RSB cavity equations
for such a model have the general form
\begin{eqnarray}
\Q_{i\to a}[\r] & = & \frac{1}{\cal Z}\int\!\!\prod_{b\in\di\backslash a}\!\! 
d\Qh_{b\to i}[\rh^{(b)}] \;\; z[\{\rh^{(b)}\};\mu_2]^{\mu_1/\mu_2}
\;\; \delta[\r-
\r^{\rm c}[\{\rh^{(b)}\};\mu_2]]\, \label{TwoStep_1}
,\\ 
\Qh_{a\to i}[\rh] & = & \int\!\!\prod_{j\in\da\backslash i}\!\! 
d\Q_{j\to\a}[\r^{(j)}]\;\;\delta[\rh-\rh^{\rm c}[\{\r^{(j)}\}]]\, ,
\label{TwoStep_2}
\end{eqnarray}
where $\mu_1$ and $\mu_2$ are the zero-temperature Parisi parameters
and satisfy the usual inequalities $0\le \mu_1\le\mu_2$. They are
related to their finite-temperature counterparts $m_1$ and $m_2$ as
follows: $\mu_i=\lim_{T\to 0} m_i(T)/T$. The integrals in
Eqs.~(\ref{TwoStep_1}), (\ref{TwoStep_2}) run over the simplex ${\cal
S}$. The delta-functions are understood to operate on the same space.

The function $\rh^{\rm c}[\r^{(1)}\dots\r^{(k-1)}]$ (here ``c'' stands
for ``cavity'') is model-dependent. We will recall in
App.~\ref{ExplicitFormulae} its precise form for the models treated in
Sec.~\ref{NumericalSection}.  The functions $\r^{\rm
c}[\rh^{(1)}\dots\rh^{(l)};\mu_2]$ and
$z[\rh^{(1)}\dots\rh^{(l)};\mu_2]$ are, on the other hand,
universal. They are defined as follows
\begin{eqnarray}
\r_q^{\rm c} = \frac{1}{z[\{\rh^{(i)}\};\mu_2]}
\sum_{(q_1\dots q_l)\in\Omega_q} \prod_{i=1}^l\;\rh^{(i)}_{q_i}\;
\ex^{-\mu_2\left(\sum_{i}|q_i|-|\sum_i q_i|\right)}\,,
\qquad q\in\{+,0,-\}\,,
\label{RhoDefinition}
\end{eqnarray}
where $\Omega_+= \{(q_1\dots q_l): \sum_i q_i>0\}$, $\Omega_0=
\{(q_1\dots q_l): \sum_i q_i=0\}$, and $\Omega_-= \{(q_1\dots q_l):
\sum_i q_i<0\}$ (with $q_i\in\{+,0,-\}$).  The normalization
$z[\{\rh^{(i)}\};\mu_2]$ is fixed by requiring $\r_+^{\rm c}+\r_0^{\rm
c}+\r_-^{\rm c}=1$.  Hereafter we will use the symbols $q, q_i,
q'\dots$ for variables running over the set $\{+,0,-\}$.

In order to select a type-II instability \cite{NostroInst}, see
Sec.~\ref{Introduction}, we shall consider order parameters which
concentrate near the ``corners'' of the simplex ${\cal S}$:
$\delta^{(+)}$, $\delta^{(0)}$, $\delta^{(-)}$.  The corner distributions
are defined by: $\delta^{(q)}_{q'} = 1$ if $q=q'$ and 0
otherwise. We can decompose such order parameters as follows:
\begin{eqnarray}
\Q_{i\to a}[\r] & = & r^{(+)}_{i\to a} \Q^{(+)}_{i\to a}[\r]
+r^{(0)}_{i\to a} \Q^{(0)}_{i\to a}[\r]
+r^{(-)}_{i\to a} \Q^{(-)}_{i\to a}[\r]\, ,\label{Decomposition_1}\\
\Qh_{a\to i}[\rh] & = & \rr^{(+)}_{a\to i} \Qh^{(+)}_{a\to i}[\rh]
+\rr^{(0)}_{i\to a} \Qh^{(0)}_{a\to i}[\rh]
+\rr^{(-)}_{i\to a} \Qh^{(-)}_{a\to i}[\rh]\, ,\label{Decomposition_2}
\end{eqnarray}
where the distributions $\Q^{(q)}_{i\to a}[\r]$ and $\Qh^{(q)}_{a\to
i}[\rh]$ are supposed to be normalized and concentrated near
$\delta^{(q)}$. We parametrize the ``width'' of these distributions
using the 6 parameters for each directed link:
\begin{eqnarray}
\e^{(q)+}_{i\to a} 
\equiv (-1)^{\delta_{q,+}}\int\!\! dQ^{(q)}_{i\to a}[\r]\,\, (\r_+ - \delta^{(q)}_+)\, ,\;\;\;\;
\e^{(q)-}_{i\to a} 
&\equiv& (-1)^{\delta_{q,-}}\int\!\! dQ^{(q)}_{i\to a}[\r]\,\, (\r_- - \delta^{(q)}_-)\, ,
\end{eqnarray}
with analogous definitions for the parameters $\eh^{(q)+}_{a\to i}$
and $\eh^{(q)-}_{a\to i}$.  Notice that the the
$(-1)^{\delta_{q,\cdot}}$ prefactors have been properly chosen to make
the $\e^{(q)\sigma}_{i\to a}$, $\eh^{(q)\sigma}_{a\to i}$ positive.

It is easy to see that the recursions (\ref{TwoStep_1}),
(\ref{TwoStep_2}) preserve the subspace $\{ \Q, \Qh$ :
$\e^{(q)\pm}_{i\to a} = \eh^{(q)\pm}_{a\to i} = 0$ for any $i$, $a$,
$q\}$. This is in fact a possible embedding of the 1RSB solution in
the 2RSB space. Of course the parameters $r^{(q)}_{i\to a}$ and
$\rr^{(q)}_{a\to i}$ must satisfy, in this case, the 1RSB equations:
\begin{eqnarray}
r_{i\to a} = \rho^{\rm c}[\{\rr_{b\to i}\}_{b\in\di\backslash a};\mu_1]
\, ,\;\;\;\;\;\;\;
\rr_{a\to i} = \rh^{\rm c}[\{r_{j\to a}\}_{j\in\da\backslash i}]\, .
\label{1RSBsolution}
\end{eqnarray}

In order to check the stability of the 1RSB subspace, we must
linearize Eqs.~(\ref{TwoStep_1}) and (\ref{TwoStep_2}) for small
$\e^{(q)\pm}_{i\to a}$, $\eh^{(q)\pm}_{a\to i}$. This yields equations
of the form
\begin{eqnarray}
\e^{(q)\sigma}_{i\to a} & \approx & \sum_{b\in\di\backslash a}\sum_{q',\sigma'}
T^{(a)}_{b\to i}(q,\sigma|q',\sigma')\; \eh^{(q')\sigma'}_{b\to i}\, ,
\label{Linear1}\\
\eh^{(q)\sigma}_{a\to i} &\approx &\sum_{j\in\da\backslash i}\sum_{q',\sigma'}
\Th^{(i)}_{j\to a}(q,\sigma|q',\sigma')\; \e^{(q')\sigma'}_{j\to a}\, ,
\label{Linear2}
\end{eqnarray}
where $\sigma,\sigma'\in\{+,-\}$. The $6\times 6$ matrices
$T^{(a)}_{b\to i}$ and $\Th^{(i)}_{j\to a}$ can be computed in terms
of the cavity functions $\r^{\rm c}[\dots]$, $\rh^{\rm c}[\dots]$ and
of the 1RSB solution, cf.\ Eq.~(\ref{1RSBsolution}). For instance,
expanding Eq.~(\ref{TwoStep_1}) we get:
\begin{eqnarray}
T^{(a)}_{b\to i}(q,\sigma|q',\sigma') =
\frac{1}{z_{q}[\{\rr_{c\to i}\};\mu_1]}
\sum_{\stackrel{\{q_c\}\in\Omega_q}{q_b=q'}}
\prod_{c\in\di\backslash a}\rr^{(q_c)}_{c\to i}\,
\ex^{-\mu_1\left(\sum_{c}|q_c|-|\sum_c q_c|\right)}\;
M^{(b)}_{\sigma,\sigma'}(\{q_c\};\mu_2)\, ,
\label{GeneralMatrix}
\end{eqnarray}
where 
\begin{eqnarray}
z_{q}[\{\rr_{c\to i}\};\mu_1] \equiv \sum_{\{q_c\}\in\Omega_q}
\prod_{c\in\di\backslash a}\rr^{(q_c)}_{c\to i}\,
\ex^{-\mu_1\left(\sum_{c}|q_c|-|\sum_c q_c|\right)}\, .
\end{eqnarray}
The matrix $M^{(b)}_{\sigma,\sigma'}(\{q_c\};\mu_2)$ is obtained by
linearizing the cavity function $\rho^{\rm
c}[\rh^{(1)}\dots\rh^{(l)};\mu_2]$, cf.\ Eq.~(\ref{RhoDefinition}),
near the corners of the simplex:
\begin{eqnarray}
M^{(i)}_{\sigma,\sigma'}(q_1\dots q_l;\mu_2) \equiv \left|
\frac{\partial \r^{\rm c}_{\sigma}}{\partial\rh^{(i)}_{\sigma'}}-
\frac{\partial \r^{\rm c}_{\sigma}}{\partial\rh^{(i)}_{0}}
\right|_{\rh^{(i)} =\delta^{(q_i)}}\, .
\end{eqnarray}
Notice that the expression (\ref{GeneralMatrix}) depends explicitly
both on $\mu_1$ and $\mu_2$, and, through $\rr^{(q)}_{a\to i}$, on
$\mu_1$.  However, it can be shown that the dependence on $\mu_2$
cancels if the stability condition is considered. We can therefore
identify $\mu_1$ with the 1RSB parameter $\mu$.

Let us now discuss how Eqs.~(\ref{Linear1}), (\ref{Linear2}) can be
used to determine whether the 1RSB solution is stable.  The idea is to
implement these recursions, together with Eq.~(\ref{1RSBsolution}),
as a message-passing algorithm \cite{RichardsonUrbanke}. One keeps in
memory the value of the 1RSB order parameter $r_{i\to a}$ (or $\rr_{a\to
i}$) and of the 6 fluctuation parameters $\e^{(q)\sigma}_{i\to a}$ (or
$\eh^{(q)\sigma}_{a\to i}$) for each directed link. At each iteration
these values are updated using Eqs.
(\ref{1RSBsolution})-(\ref{Linear2}) for all the links of the graph
(one should imagine the old values to be used on the right hand side
and the new ones coming out on the left hand side).  The matrices
$T^{(a)}_{b\to i}$ and $\Th^{(i)}_{j\to a}$ must be recomputed after
each sweep in terms of the most recent values of $r_{i\to a}$ and
$\rr_{a\to i}$. After a fast transient the numbers $r_{i\to a}$ and
$\rr_{a\to i}$ converge to the 1RSB solution.  As for the
$\e^{(q)\sigma}_{i\to a}$ and $\eh^{(q)\sigma}_{a\to i}$, either they
converge to $0$ or they stay different from $0$ (and, in fact,
diverge).  In the first case the 1RSB solution is stable, in the
second one it is unstable.

The above procedure can be improved in several aspects. First of all
one can define (and monitor) an appropriate norm of the $\e$'s, e.g.
\begin{eqnarray}
\| \e \| = \frac{1}{N}\sum_{i=1}^N\sum_{a\in\di}\sum_{q,\sigma}
|\e^{(q)\sigma}_{i\to a}|\, .
\end{eqnarray}
After each updating sweep one can renormalize the $\e$'s by setting
$\lambda = \| \e \|$ and $\e^{(q)\sigma}_{i\to a}\leftarrow
\e^{(q)\sigma}_{i\to a}/\lambda$.  The 1RSB solution is stable if
$\lambda<1$, and unstable otherwise.  It is evident that $\lambda$
converges to the largest eigenvalue of the linear transformation
(\ref{Linear1}), (\ref{Linear2}).  The determination of the
``stability parameter'' $\lambda$ can be improved by averaging it over
many iterations of the algorithm.

In the next Sections we will be interested in evaluating the stability
threshold for {\it ensembles} of models. In this case one can
implement the same algorithm as before drawing the local structure of
the graph randomly at each iteration
\cite{MezardParisiBethe}. Moreover, by cleverly exploiting the
structure of the model to be studied, one can reduce the number of
fluctuation parameters $\e$, $\eh$ per link. In both the examples to
be studied below, it is possible to use just one parameter $\e$ and
one $\eh$ per link.

As a final remark, let us notice that the method outlined in this
Section for $T=0$, can be generalized to finite-temperature
calculations \cite{NostroFiniteT}.
%
%
\section{Numerical evaluation of the stability condition}
\label{NumericalSection}

In this Section we treat two {\it ensembles} of satisfiability models
having Hamiltonians of the form (\ref{Hamiltonian}): random $k$-XORSAT
and random $k$-SAT. In both cases the $k$-uple of sites $(i^a_1\dots
i^a_k)$ involved in a given clause $a$ is chosen with flat probability
distribution among the $\binom{N}{k}$ possible $k$-uples.

The zero-temperature phase diagram of these models (for $k \ge 3$) is
known \cite{3SAT,Variational,XorSat,MarcGiorgioRiccardo,MarcRiccardo,
CoccoEtAlXorSat,MezardEtAlXorSat} to be composed by 3 different phases
as a function of $\alpha$, cf.\ Figs.~\ref{StabilityXORFig} and
\ref{StabilitySATFig}.  For $\alpha < \alpha_{\rm d}$ the system is
paramagnetic with no diverging energy barriers in the configurational
space. For $\alpha_{\rm d} \le \alpha \le \alpha_{\rm c}$ the model is
still unfrustrated: the ground state energy is zero.  Nevertheless the
Gibbs measure decomposes in an exponentially large number of pure
states separated by large energy barriers.  For $\alpha > \alpha_c$
frustration percolates and the ground state energy becomes
positive. The ground state is still hidden within a large number of
metastable states.

Within the 1RSB approximation, the system is completely described by
the complexity $\Sigma(e)$, i.e.\ the normalized logarithm of the
number of metastable states with energy density $e$. For
$\alpha>\alpha_{\rm d}$, the complexity becomes strictly positive in
the interval $e \in [e_{\rm s}(\alpha),e_{\rm d}(\alpha)]$.  The
static energy $e_{\rm s}(\alpha)$ vanishes for $\alpha \le \alpha_{\rm
c}$ and becomes positive above the static transition point
$\alpha_{\rm c}$.  The 1RSB dynamical energy $e_{\rm d}$, becomes
positive at the dynamical critical point $\alpha_{\rm d}$.

The 1RSB calculation of $\Sigma(e)$ becomes quite generally
\cite{NostroInst} unstable with respect to further replica-symmetry
breakings above the Gardner energy $e_{\rm G}$, with $e_{\rm G}<e_{\rm
d}$. We expect the 1RSB result to be correct only for $e \le e_{\rm
G}$.  In the following we shall present our results for $e_{\rm
G}(\alpha)$ obtained with the method outlined in the previous Section.
%
%
\subsection{$k$-XORSAT}
\label{NumericalXORSAT}

The definitions given so far are completed by taking
\begin{eqnarray}
E_a(\us_{\da}) = 1-J_a\,\sigma_{i^a_1}\cdots\sigma_{i^a_k}\, ,
\end{eqnarray}
where the $J_a$'s are i.i.d.\ (quenched) random variables taking the
values $\pm 1$ with equal probabilities. We recall
\cite{Creignou,XorSat} that the existence of an unfrustrated ground
state can be mapped onto the existence of a solution for a certain
random linear system over $G{\mathbb F}[2]$.

\begin{figure}
\begin{center}
\includegraphics[width=0.65\textwidth]{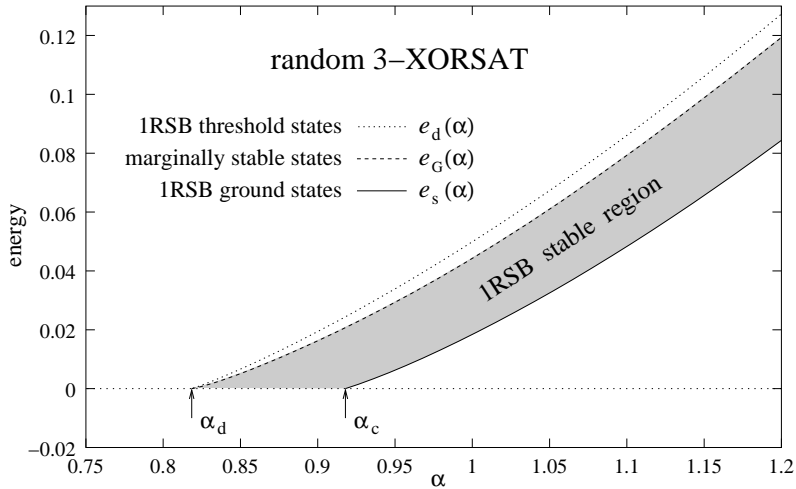}
\end{center}
\caption{The stability region in the energy-clause density plane for
3-XORSAT. We used population dynamics algorithms with $O(10^5)$
elements and $O(10^2)$ iterations. The marginal stability line $e_{\rm
G}(\alpha)$ crosses the one-step ground state energy $e_{\rm
s}(\alpha)$ at $\alpha_{\rm G}=3.072(2) $ (outside the range shown
here).}
\label{StabilityXORFig}
\end{figure}
In Fig.~\ref{StabilityXORFig} we present our results for the zero
temperature phase diagram for $k=3$ (but the picture remains
qualitatively similar for any $k$).  For $k=3$ we have $\alpha_{\rm d}
\approx 0.81846916$ and $\alpha_{\rm c} \approx 0.91793528$.  It is
evident from the data of Fig.~\ref{StabilityXORFig} that $0<e_{\rm
G}(\alpha)<e_{\rm d}(\alpha)$ for $\alpha >\alpha_{\rm d}$, and that
$e_{\rm G}(\alpha), e_{\rm d}(\alpha)\downarrow 0$ as
$\alpha\downarrow\alpha_{\rm d}$. This is confirmed by a perturbative
expansion for $\alpha\to\alpha_{\rm d}$, cf.\
App.~\ref{ExpansionXORSAT}.  In particular we get
\begin{eqnarray}
e_{\rm d}(\alpha) = e_{\rm d}^{(0)}(\alpha-\alpha_{\rm d})^{1/2} +
O(\alpha-\alpha_{\rm d})\, ,\;\;\;\;\;\;
e_{\rm G}(\alpha) = e_{\rm G}^{(0)}(\alpha-\alpha_{\rm d}) +
O((\alpha-\alpha_{\rm d})^{3/2})\, ,\label{ExpansionMainText}
\end{eqnarray}
with $e_{\rm d}^{(0)} \approx 0.0107506548$ and $e_{\rm G}^{(0)}
\approx 0.0753987711$ for $k=3$.  Let us stress that our result
$e_{\rm G}(\alpha)>0$ for $\alpha>\alpha_{\rm d}$ is consistent with
the rigorous solution of the model at $e=0$
\cite{CoccoEtAlXorSat,MezardEtAlXorSat}.

In the limit $\alpha\to\infty$ we expect to recover the fully
connected $k$-spin Ising spin glass. At zero temperature this model is
characterized by FRSB even for what concerns the ground state
concerns~\cite{Gardner}. This implies that the line $e_{\rm
G}(\alpha)$ must cross $e_{\rm s}(\alpha)$ at some finite $\alpha_{\rm
G}$. This expectation is in fact fulfilled by our data: we get
$\alpha_{\rm G} =3.072(2)$ for $k=3$. For $\alpha>\alpha_{\rm G}$ the
1RSB result for the ground state energy is just a lower bound
\cite{FranzLeone}.
%
%
\subsection{$k$-SAT}
\label{NumericalKSAT}

In this case the energy of a clause is 
\begin{eqnarray}
E_a(\us_{\da}) = 2\prod_{i\in\da}\frac{1-J_{a\to i}\sigma_i}{2}\, ,
\end{eqnarray}
where the $J_{a\to i}$ are i.i.d.\ (quenched) random variables taking
the values $\pm 1$ with equal probabilities. The clause has energy 2
(it is not satisfied) if all of participating spins $\sigma_i$ have
opposite signs to the corresponding $J_{a\to i}$. In all the other
cases it has energy 0 (it is satisfied).
\begin{figure}
\begin{tabular}{c}
\includegraphics[width=0.65\textwidth]{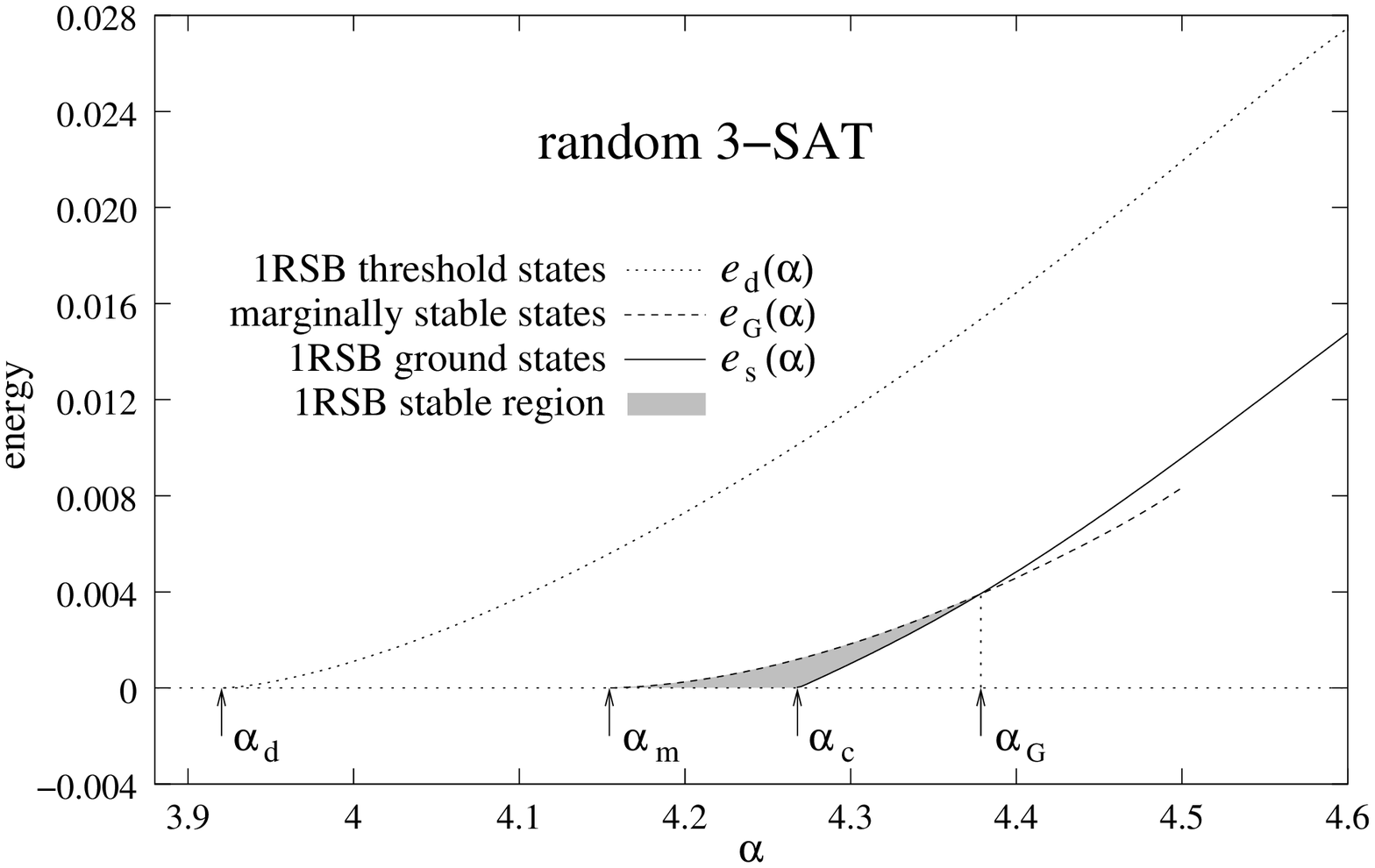}\\
\includegraphics[width=0.65\textwidth]{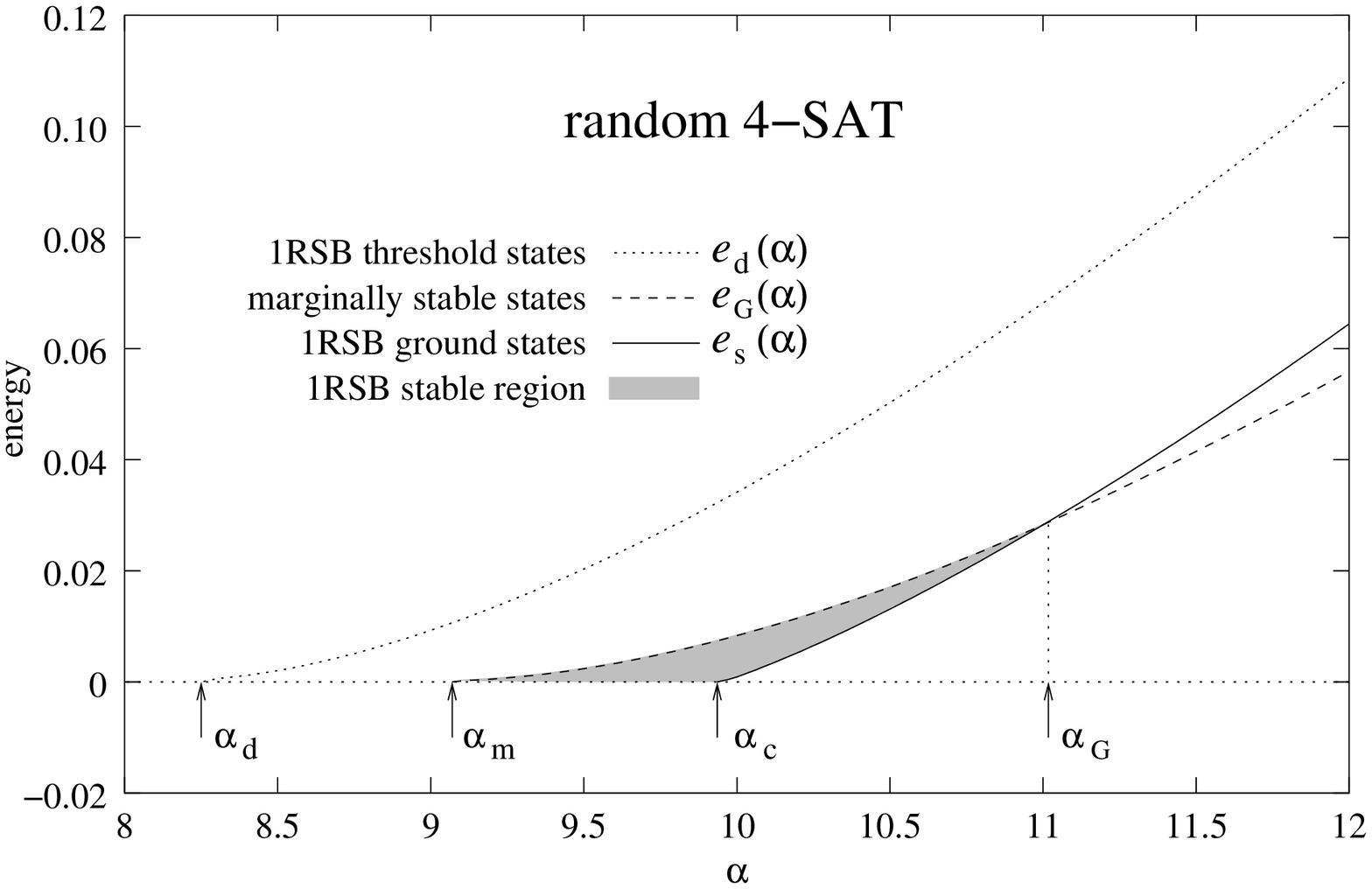}
\end{tabular}
\caption{The stability region in the energy-clause density plane for
3-SAT (top) 4-SAT (bottom). Here we used population dynamics
algorithms with (typically) $O(10^5)$ elements and $O(10^2)$
iterations.}
\label{StabilitySATFig}
\end{figure}
In Fig.~\ref{StabilitySATFig} we show the results of our method for
the stability threshold $e_{\rm G}(\alpha)$, together with the curves
for $e_{\rm s}(\alpha)$ and $e_{\rm d}(\alpha)$ computed in
Refs. \cite{MarcGiorgioRiccardo,MarcRiccardo}.  We considered the two
prototypical cases $k=3$ and $k=4$, but we expect the picture to
remain qualitatively similar for any $k\ge 3$.  There is an important
qualitative difference with respect to $k$-XORSAT: for $k$-SAT $e_{\rm
G}(\alpha)\downarrow 0$ for $\alpha\downarrow \alpha_{\rm m}$ with
$\alpha_{\rm d}< \alpha_{\rm m} <\alpha_{\rm c}$. In other words even
zero-energy states become unstable towards FRSB at sufficiently low
$\alpha$.  In the 1RSB picture, zero-energy states are related to
cluster of solutions of the corresponding satisfiability problem
\cite{Variational,MarcGiorgioRiccardo,MarcRiccardo}. It would be
interesting to understand how this picture must be modified below
$\alpha_{\rm m}$. Generalizing the ideas of \cite{NostroInst}, we
expect, within a 2RSB description, such clusters to split continuously
into sub-clusters at $\alpha_{\rm m}$.
\begin{figure}
\begin{center}
\includegraphics[width=0.65\textwidth]{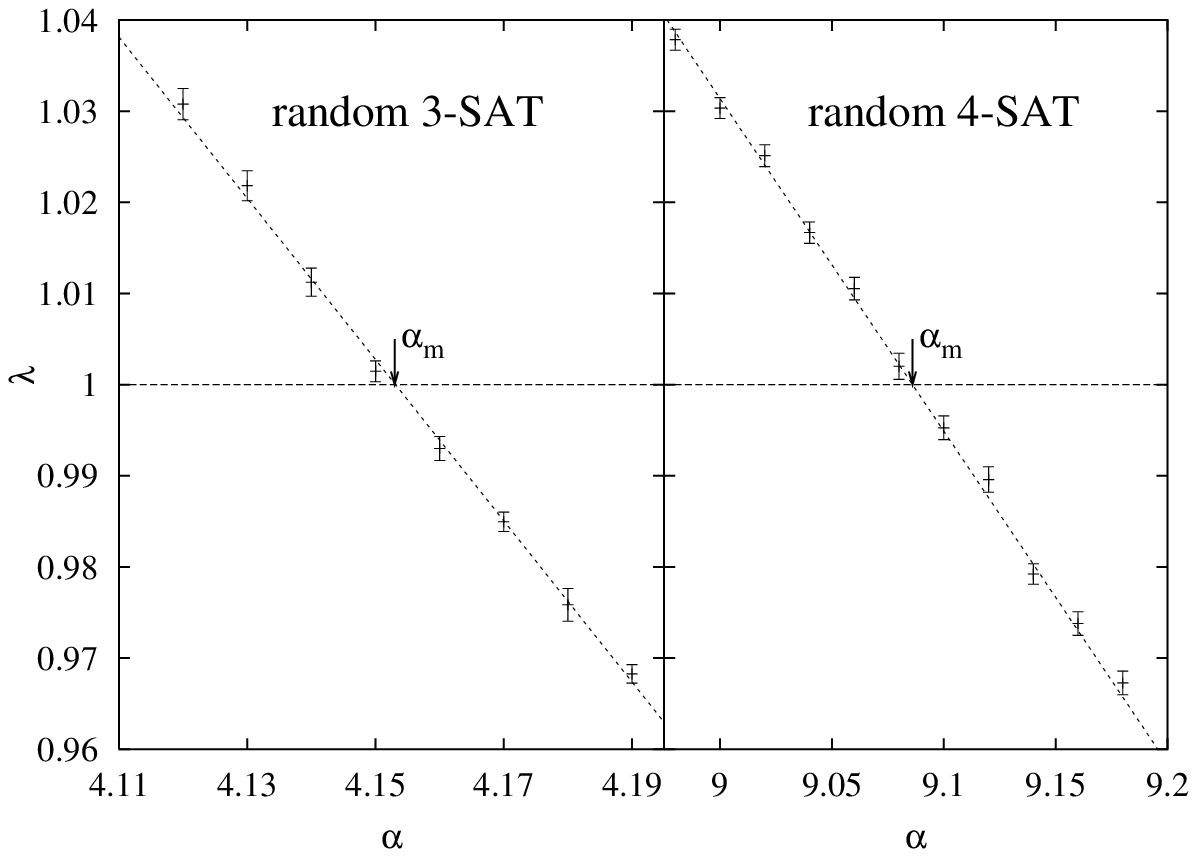}
\end{center}
\caption{The stability parameter (largest eigenvalue of the stability
matrix) $\lambda(\alpha,e=0)$ for 3-SAT (left) and 4-SAT (right).}
\label{LambdaFig}
\end{figure}

Unlike $k$-XORSAT, the stability computation has a non-trivial result
even at zero-energy. It is therefore interesting to modify the
approach of Sec.~\ref{GeneralSection} in order to consider this limit
case.  The zero-energy limit of the 2RSB equations (\ref{TwoStep_1}),
(\ref{TwoStep_2}) is obtained by taking $\mu_1,\mu_2\to\infty$ with
$\mu_1/\mu_2=x$ fixed. As already noticed in
Sec.~\ref{GeneralSection}, the stability parameter $\lambda$ ends up
not depending upon $\mu_2$, and therefore $x$. We report our results
for $\lambda(\alpha, e=0)$ in Fig.~\ref{LambdaFig}.  This approach
provide us good estimates of $\alpha_{\rm m}$.  We get $\alpha_{\rm m}
= 4.153(1)$ for $k=3$ and $\alpha_{\rm m} = 9.086(2)$ for $k=4$. This
should be compared with the values $\alpha_{\rm d} = 3.925(3)$ and
$\alpha_{\rm c} = 4.266(1)$ for $k=3$, and $\alpha_{\rm d} = 8.295(6)$
and $\alpha_{\rm c} = 9.931(2)$ for $k=4$.

Finally, in the large connectivity limit, the 1RSB solution is
unstable even for the ground state. The curves $e_{\rm
G}(\alpha)$ and $e_{\rm s}(\alpha)$ cross at $\alpha_{\rm G}$.  We get
$\alpha_{\rm G} = 4.390(5)$ for $k=3$ and $\alpha_{\rm G} = 11.00(2)$
for $k=4$.

%
%
\section{On the zero-temperature limit of the 1RSB solution}
\label{NonIntegerSection}

In this Section we consider the finite-temperature 1RSB solution and
discuss its $T\to 0$ limit.  At finite temperature the 1RSB order
parameter is given by a probability distribution over the reals for
each directed link.  We shall call these distributions $\rho_{i\to
a}(h)$ and $\rh_{a\to i}(u)$. The variables $u$ and $h$ are ``cavity
fields''.  If the factor graph were a tree, we could define the cavity
fields as follows.  Consider the branch $a\to i$ of the graph: this is
the connected sub-tree which has i as the root and contains only $a$
among the neighbors of $i$.  Let $Z_{a\to i}(\sigma_i)$ be the
partition function for this subsystem constrained to a given value of
$\sigma_i$, at inverse temperature $\beta$.  This quantity can be
parametrized in terms of a cavity field $u_{a\to i}$ as follows
\begin{eqnarray}
Z_{a\to i}(\sigma_i) = Z^{(0)}_{a\to i}\, e^{\beta u_{\a\to i}
\sigma_i}\, .
\end{eqnarray}
It is now elementary to show that, if the Hamiltonian has the form
(\ref{Hamiltonian}), with $E_a(\us_{\da})$ taking values in $\{0,2\}$,
the field $u_{\a\to i}$ must become an integer number when $T\to 0$.
The same conclusion is easily reached for the fields $h_{i\to a}$.

The situation is less clear on locally tree-like graphs, such as the
ones considered in Sec. \ref{NumericalSection}.  However, one can
argue that the same property of the $T\to 0$ limit must hold {\it
within} each pure state.  Suppose now that the 1RSB cavity fields
distributions $\rho_{i\to a}(h)$ or $\rh_{a\to i}(u)$ have a
non-vanishing support over non-integer fields even in the $T\to 0$
limit.  It is reasonable to take this as an evidence for the 1RSB
ansatz being incorrect (this kind of argument was pioneered in Ref.
\cite{MonassonRSB}).  In fact we do not expect the properties of the
system to be discontinuous at $T=0$.

Let us notice in passing that other interesting phenomena appear when
$T>0$.  At infinitesimal temperature the cavity fields acquire a small
part proportional to the temperature. These evanescent contributions
can in turn undergo one or several replica-symmetry-breakings
transitions \cite{Variational}. This is what happens in 3-SAT above
$\alpha_{\rm b}\approx 3.87<\alpha_{\rm d}$ \cite{GPSAT}.  The
underlying physical phenomenon is the following.  In the region
$\alpha_{\rm b}< \alpha<\alpha_{\rm d}$ the space of solutions
decomposes into clusters.  However these cluster do not have any
backbone (i.e. a subset of the variables which is fixed in all the
solution of the cluster).  The backbone percolates at $\alpha_{\rm
d}$.  In this Section we focus on ``hard'' cavity fields (i.e.\
non-vanishing in the zero-temperature limit) and these effects do not
concern us.  Notice that ``hard'' fields are the ones determining the
energy in the $T\to 0$ limit.

It is easy to understand why a cavity fields distribution not
concentrated on the integers in the $T\to 0$ limit, should be related
to the instability of the 1RSB solution.  Indeed the effect of the
finite temperature is to provide fields that are not integers
but differ from integers by a term of order $T$.  When we insert these
slightly-non-integer fields in the cavity equations this
perturbation may be amplified, and after a finite number of steps the
distribution may spread over non-integer fields. This instability
corresponds to the propagation of a perturbation at any distance and it
is a signal of instability of the 1RSB solution.

We investigated this phenomenon analytically for $k$-XORSAT and
$k$-SAT. The idea is to write the 1RSB order parameter as follows
\begin{eqnarray}
\rh_{a\to i}(u) = \rr^{(+)}_{a\to i}\delta(u-1) + \rr^{(0)}_{a\to i}\delta(u) +
\rr^{(-)}_{a\to i}\delta(u+1) + \eh^{(+)}_{a\to i}(u) + 
\eh^{(-)}_{a\to i}(u)\, ,
\end{eqnarray}
and analogously for $\rho_{i\to a}(h)$, with $\eh^{(+)}_{a\to i}(u)$,
$\eh^{(-)}_{a\to i}(u)$ small perturbations supported, respectively
over $(0,1)$, and $(-1,0)$ (it turns out that $|u|\le 1$ always).  
We then considered the $T=0$ cavity equations with 1RSB parameter
$\mu$. It is easy to realize that the 1RSB equations  
leave the $\eh_{a\to i}, \e_{i\to a}= 0$
subspace invariant (we worked indeed within this subspace in the rest
of the paper). The next step is therefore to expand
for small perturbations $\e^{(\pm)}_{i\to a}(h)$
and $\eh^{(\pm)}_{a\to i}(u)$.
The resulting linear equations have a simple invariant subspace:
\begin{eqnarray}
\eh^{(\pm)}_{a\to i}(u)=\deh^{(\pm)}_{a\to i}\, \ex^{\mu |u|}\, ,
\label{FiniteTsubspace}
\end{eqnarray}
and analogously for $\e^{(\pm)}_{i\to a}(h)$ (with parameters
$\delta^{(\pm)}_{i\to a}$). One can therefore deduce a set of linear
recursions for the parameters $\deh^{(\pm)}_{a\to i}$,
$\delta^{(\pm)}_{i\to a}$. It turns out that, both for
$k$-SAT and for $k$-XORSAT, these recursions are equivalent to 
the ones obtained in the previous Sections for the 2RSB perturbations.
Under the assumption that Eq. (\ref{FiniteTsubspace}) defines the most relevant
(unstable) subspace, this implies that stability
with respect to non-integer fields is indeed equivalent
to stability with respect to 2RSB perturbations.

\begin{figure}
\begin{center}
\includegraphics[width=0.65\textwidth]{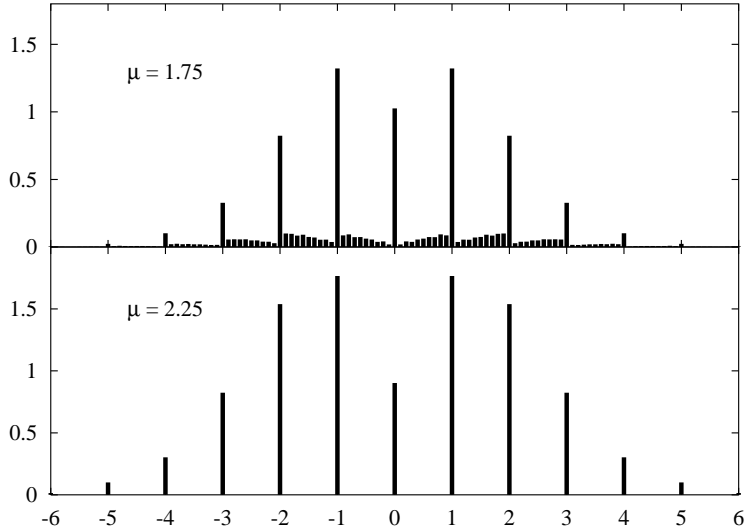}
\end{center}
\caption{Cavity field distribution $\rho_{i\to a}(h)$ of 3-SAT,
averaged over all the directed links of the factor graph.  Here we
used $\mu=1.75$ (upper frame) and $2.25$ (ower frame), and
$\alpha=4.51$. The 1RSB equations were solved by a population dynamics
algorithm with $4000$ populations of $256$ fields each.}
\label{FieldsFig}
\end{figure}
For finite $\mu$, we analyzed this instability numerically.  As an
example, in Fig.~\ref{FieldsFig} we report the results of a numerical
solution of the 1RSB equations for 3-SAT at $\alpha= 4.51>\alpha_{\rm G}$.  
Here we used $T=10^{-5}$ and 1RSB parameter $\mu =1.75 $ and $2.25 $, which
should be confronted with the ground-state value $\mu_{\rm
s}(\a=4.51)= 1.94(1)$.  The first choice of $\mu$ corresponds to
metastable states, while the second to an energy below the ground
state.  It is clear form the figure that $\mu=1.75$ is in the unstable
region while $\mu=2.25$ is in the stable region.

In fact the distributions $\rho_{i\to a}(h)$ or $\rh_{a\to i}(u)$
acquire a non-vanishing support on non-integer fields as soon as $\mu
< \mu_{\rm int}(\alpha)$ (i.e. for energy $e>e_{\rm int}(\alpha)$).  A
precise numerical determination of $\mu_{\rm int}(\alpha)$
is not simple because the solution of the cavity
equations at finite temperature is a rather slow process.  Moreover
many systematic effects must be taken into account.  As mentioned
above consistency implies that $\mu_{\rm int}(\alpha)\le\mu_{\rm
G}(\alpha)$. Furthermore, we argued $\mu_{\rm int}(\alpha)=\mu_{\rm
G}(\alpha)$under the assumption that 
Eq. (\ref{FiniteTsubspace}) defines indeed the most relevant
perturbation. 
Our numerical best estimates give $\mu_{\rm
int}(4.51)=2.00(3)$. This is numerically compatible with the result
obtained with the methods of Secs.~\ref{GeneralSection} and
\ref{NumericalSection}: $\mu_{\rm G}(4.51)= 2.045(5)$.

\begin{figure}
\begin{center}
\includegraphics[width=0.65\textwidth]{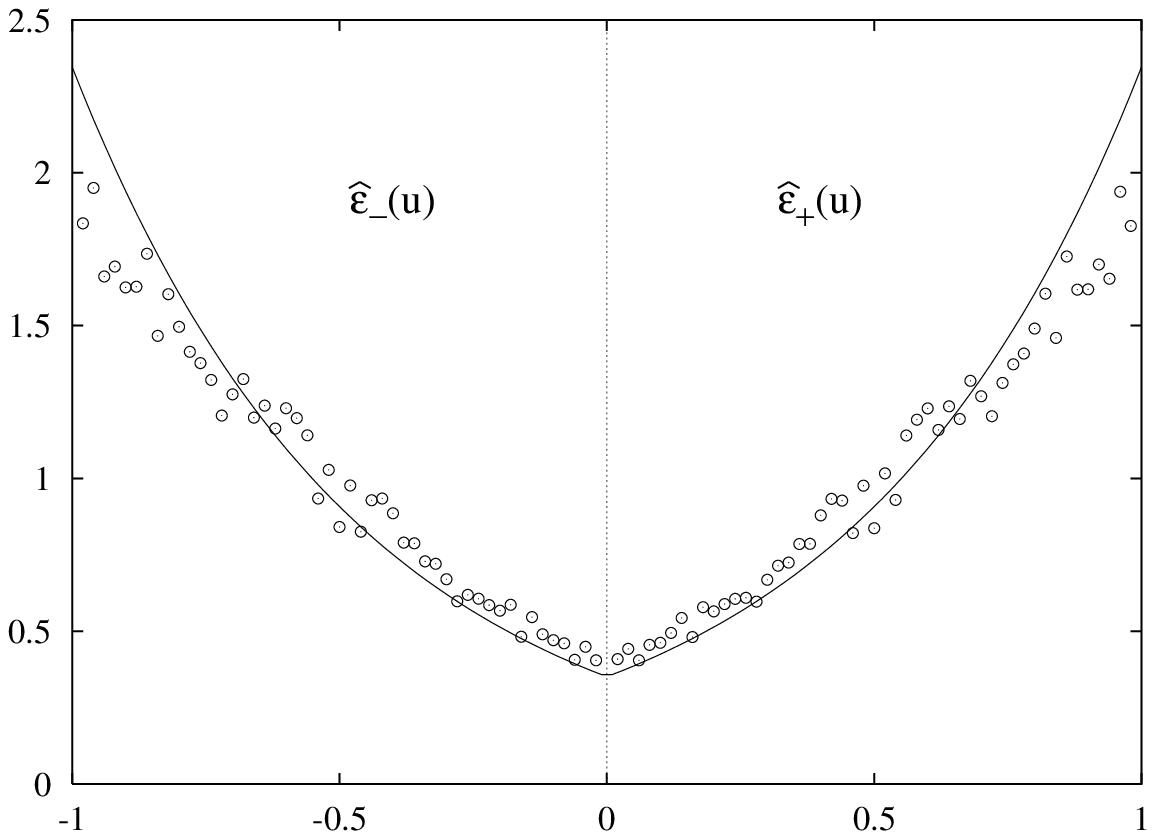}
\end{center}
\caption{Distributions $\eh^{(\pm)}_{a\to i}(u)$ of non-integer cavity fields for 3-SAT,
averaged over all the directed links of the factor graph.  Here we 
used $\alpha=4.51$ and $\mu=1.9<\mu_{\rm int}$ and normalized the integral of $\eh^{(\pm)}_{a\to i}(u)$
to 1. The continuous line is a fit to the expectated form $\deh_* \exp(\mu|u|)$,
cf. \ref{FiniteTsubspace}. We found the value $\deh_* \approx 0.351$ for the best fitting coefficient.}
\label{UFieldsFig}
\end{figure}
Finally in Fig. \ref{UFieldsFig} we present our numerical data for 
the average of the 
distributions $\rh_{a\to i}(u)$ at $\alpha=4.51$
and $\mu = 1.9<\mu_{\rm int}(\alpha)$. 
If the analytical argument outlined above is
correct and in the approximation that $|\mu_{\rm int}-\mu|\approx 0.10\ll 1$, 
we should have  $\overline{\eh^{(\pm)}_{a\to i}(u)}\approx \deh_* \exp(\mu|u|)$. 
The resonable agreement confirms that
Eq. (\ref{FiniteTsubspace}) corresponds to 
the most relevant subspace for finite-$T$ perturbations.
%
%
\section{Conclusion}
\label{Conclusions}

Our main conclusion is that FRSB plays an important role in random
combinatorial optimization problems such as $k$-XORSAT and $k$-SAT. We
investigated this issue by analyzing the stability of the cavity
recursions, both within a 2RSB ansatz, cf.\ Secs.~\ref{GeneralSection}
and \ref{NumericalSection}, and at finite temperature, cf.\
Sec.~\ref{NonIntegerSection}.

The 1RSB ground state becomes unstable in two different regimes.  In
the UNSAT region, it becomes unstable in highly constrained problems:
$\alpha>\alpha_{\rm G}$, i.e. when the corresponding factor graph has
large connectivity.  This result was not unexpected.  As shown in
Refs. \cite{Gardner} (for $k$-XORSAT) and
\cite{LeuzziParisi,CrisantiLeuzzi} (for 3-SAT), in the $\alpha \to
\infty$ limit, these models have a low temperature FRSB phase.  In the
SAT region there is an exponential number of unfrustrated ground
states (i.e. solutions of the satisfiability problem).  The 1RSB
solution indicates that these solutions have a clustered structure in
the region $\alpha_{\rm d}<\alpha<\alpha_{\rm c}$.  However, this
solution becomes unstable in the underconstrained region $\alpha_{\rm
d}<\alpha<\alpha_{\rm m}$. Providing a more refined description of the
space of solutions in this regime is an open problem (which will
require presumably FRSB).

Let us recall that high-lying metastable states are {\it always}
unstable against FRSB. This could have remarkable consequences on the
performances of local search algorithms (such as simulated
annealing). Let us suppose, just to estimate how large this effect can
be, that there are no metastable states above the instability energy
$e_{\rm G}(\alpha)$. This would imply that the total number of
metastable states at the SAT-UNSAT phase transition is $\Sigma(e_{\rm
G}(\alpha_{\rm c}),\alpha_{\rm c}) \approx 0.0019$ (for 3-SAT) and
$0.0099$ (for 4-SAT), instead of $\Sigma(e_{\rm d}(\alpha_{\rm
c}),\alpha_{\rm c})\approx 0.010$ (for 3-SAT) and $0.029$ (for
4-SAT). In other words metastability would start having some effect
only at much larger sizes.

In this work we studied the stability of the 1RSB phase, by analyzing
the stability of the cavity recursions with respect to two types of
perturbations.  In Secs.~\ref{GeneralSection} and
\ref{NumericalSection} we considered a perturbation towards 2RSB,
while in Sec.~\ref{NonIntegerSection} we used a perturbation towards
non-integer fields. We argued that the two approaches give indeed
coincident answers, cf. Sec.~\ref{NonIntegerSection}.
This leaves open the questions whether there can
be more dangerous instabilities. In replica formalism, one should
diagonalize the Hessian of the free energy in the full replica
space. We expect that our calculation corresponds to selecting one
particular subspace. The question is whether this is the most
dangerous subspace. The fact \cite{NostroFiniteT} that in the
$\alpha\to\infty$ limit our calculation yields the replicon
instability points towards a positive answer
%
%
\begin{acknowledgments}
FRT thanks ICTP for kind hospitality during the completion of this
manuscript.  Work supported in part by the European Community's Human
Potential Programme under contract HPRN-CT-2002-00307, Dyglagemem.
\end{acknowledgments}

%
%
\appendix
\section{Explicit formulae}
\label{ExplicitFormulae}

In this Appendix we give the explicit forms of the recursions
(\ref{1RSBsolution}), (\ref{Linear1}) and (\ref{Linear2}) for the
models treated in Sec.~\ref{NumericalSection}. Moreover we show how to
reduce the size of the $6\times 6$ matrices $T^{(a)}_{b\to i}$ and
$\Th^{(i)}_{j\to a}$ using the symmetries of the models. Finally, it
will become evident that the stability parameter $\lambda$ depends
uniquely upon the smallest replica-symmetry breaking parameter
$\mu_1$.

Before turning to the models of Sec.~\ref{NumericalSection}, it is
useful to compute the matrix $M^{(i)}_{\sigma,\sigma'}(q_1\dots
q_l;\mu_2)$, cf.\ Eq.~(\ref{GeneralMatrix}), which is
model-independent.  We get the following result
\begin{eqnarray}
M^{(i)}_{\sigma,\sigma'}(q_1\dots q_l;\mu_2)=\left\{
\begin{array}{cl}
0\;\;& \mbox{   if $\sum_{j\neq i}q_j >1$ or $<1$,}\\
\ex^{-2\mu_2}\delta_{\sigma,+}\delta_{\sigma',-}\;\;& 
\mbox{   if $\sum_{j\neq i}q_j =1$ and $q_i=0,+$,}\\
\ex^{2\mu_2}\delta_{\sigma,+}\delta_{\sigma',-}\;\;&
\mbox{   if $\sum_{j\neq i}q_j =1$ and $q_i=-$,}\\
\ex^{-2\mu_2}\delta_{\sigma,-}\delta_{\sigma',+}\;\;&
\mbox{   if $\sum_{j\neq i}q_j =-1$ and $q_i=0,-$,}\\
\ex^{2\mu_2}\delta_{\sigma,-}\delta_{\sigma',+}\;\;&
\mbox{   if $\sum_{j\neq i}q_j =-1$ and $q_i=+$,}\\
\delta_{\sigma,\sigma'}\;\;&
\mbox{   if $\sum_{j\neq i}q_j =0$.}
\end{array}\right.\label{M_matrix}
\end{eqnarray}
%
%
%
\subsection{$k$-XORSAT}
\label{ExplicitXORSAT}

Let us start by recalling that the function 
$\rh^{\rm c}[\r^{(1)}\dots \r^{(k-1)}]$ takes in this case the form:
\begin{eqnarray}
\rh^{\rm c}_+ & = & \frac{1}{2} \left[\prod_{i=1}^{k-1}(\r^{(i)}_++\r^{(i)}_-)+
\prod_{i=1}^{k-1}(\r^{(i)}_+-\r^{(i)}_-)\right]\, ,\\
\rh^{\rm c}_0 & = & 1-\prod_{i=1}^{k-1}(\r^{(i)}_++\r^{(i)}_-)\, \\
\rh^{\rm c}_- & = & \frac{1}{2} \left[\prod_{i=1}^{k-1}(\r^{(i)}_++\r^{(i)}_-)-
\prod_{i=1}^{k-1}(\r^{(i)}_+-\r^{(i)}_-)\right]\, .
\end{eqnarray}
This completely specifies the 2RSB saddle point equations
(\ref{TwoStep_1}), (\ref{TwoStep_2}). Moreover, it turns out that the
1RSB solution is symmetric under spin inversion: $r^{(+)}_{i\to
a}=r^{(-)}_{i\to a}$, and $\rr^{(+)}_{a\to i}=\rr^{(-)}_{a\to i}$
\cite{MezardEtAlXorSat}.  We can therefore parametrize it in terms of
a single real number per directed link, e.g.\ $r^{(0)}_{i\to a}$,
$\rr^{(0)}_{a\to i}$. Using this simplifying features it is easy to
show that the matrices $T^{(a)}_{b\to i}$, $\Th^{(i)}_{j\to a}$ have
the following form:
\begin{eqnarray}
T^{(a)}_{b\to i} = \left[ \begin{array}{cccccc}
C_1 & C_2 & 0 & C_3 & 0 & 0\\
0 & C_1 & 0 & 0 & 0 & 0\\
0 & 0 & C_4 & 0 & 0 & C_5\\
C_5 & 0 & 0 & C_4 & 0 & 0\\
0 & 0 & 0 & 0 & C_1 & 0\\
0 & 0 & C_3 & 0 & C_2 & C_1
\end{array}
\right]\, ,\;\;\;\;\;\;
\Th^{(i)}_{j\to a} = \left[ \begin{array}{cccccc}
1/2 & 0 & 0 & 0 & 0 & 1/2\\
0 & 1/2 & 0 & 0 & 1/2 & 0\\
0 & 0 & \Ch & \Ch & 0 & 0\\
0 & 0 & \Ch & \Ch & 0 & 0\\
0 & 1/2 & 0 & 0 & 1/2 & 0\\
1/2 & 0 & 0 & 0 & 0 & 1/2
\end{array}
\right]\, ,
\end{eqnarray}
where we ordered the 6 components as follows 
$(q,\sigma)= \{ (+,+),(+,-);$ $(0,+),(0,-);$ $(-,+),$ $(-,-)\}$.
The constants $C_1,\dots, C_5$ are easily computed using Eqs. 
(\ref{GeneralMatrix}) and (\ref{M_matrix}), while 
$\Ch=\Ch^{(i)}_{j\to a}$ is given by
\begin{eqnarray}
\Ch^{(i)}_{j\to a} = \frac{1}{2}\;
\frac{r^{(0)}_{j\to a}\prod_{l\in\da\backslash \{i,j\}}
(1-r^{(0)}_{l\to a})}{1-\prod_{l\in\da}(1-r^{(0)}_{l\to a})}\, .
\end{eqnarray}

As we already pointed out, the 6-dimensional transformation defined
above can be reduced thanks to the symmetries of the system. Consider
indeed the following parametrization of the fluctuation variables in
terms of the numbers $\de_{i\to a}$:
\begin{eqnarray}
\e^{(+)+}_{i\to a} = (\ex^{\mu_1-2\mu_2}/r^{(+)}_{i\to a})\,\de_{i\to a}
\, ,&&
\e^{(+)-}_{i\to a} = 0\, ,\\
\e^{(0)+}_{i\to a} = (1/r^{(0)}_{i\to a})\, \de_{i\to a}\, ,
\;\;\;\;\;\;\;\;\;\,&&
\e^{(0)-}_{i\to a} = (1/r^{(0)}_{i\to a})\, \de_{i\to a}\, .\\
\e^{(-)+}_{i\to a} = 0\, ,\;\;\;\;\;\;\;\;\;\;\;\;\;\;\;\;\;\;
\;\;\;\;\;\;\;\;\,&&
\e^{(-)-}_{i\to a} = (\ex^{\mu_1-2\mu_2}/r^{(-)}_{i\to a})\,\de_{i\to a}\, ,
\end{eqnarray}
and the analogous for $\eh^{(q)\sigma}_{a\to i}$ (in terms of $\deh_{i\to a}$).
It is easy to show that the linear subspace defined in this way is preserved 
by the transformations (\ref{Linear1}), (\ref{Linear2}). Moreover, a 
numerical calculation confirms that the largest eigenvalue $\lambda$
belongs to this subspace. Therefore, instead of 
Eqs. (\ref{Linear1}), (\ref{Linear2}) we can iterate the simpler recursion:
\begin{eqnarray}
\de_{i\to a} = \sum_{b\in\di\backslash a} t^{(a)}_{b\to i}\deh_{b\to i}\, ,
\;\;\;\;\;\;\;
\deh_{a\to i} = \sum_{j\in\da\backslash i} \th^{(i)}_{j\to a} \de_{j\to a}\, .
\label{ReducedRecursion}
\end{eqnarray}
The real numbers $t^{(a)}_{b\to i}$ and $\th^{(i)}_{j\to a}$ can be
derived from the matrices $T^{(a)}_{b\to i}$ and $\Th^{(i)}_{j\to a}$.
The result is
\begin{eqnarray}
t^{(a)}_{b\to i} & = &\frac{1}{z[\{\rr_{c\to i};\mu_1]}
\sum_{\{q_c\}}{}' \prod_{c\in\di\backslash\{a,b\}}
(\rr^{(q_c)}_{c\to i} \ex^{-\mu_1|q_c|})\label{Coefficients1}\\
\th^{(i)}_{j\to a}&=& \prod_{l\in\da\backslash\{i,j\}}(1-r^{(0)}_{l\to a})\, ,
\label{Coefficients2}
\end{eqnarray}
where the sum $\sum'$ is over the $\{ q_c\}$ such that 
$\sum_{c\in\di\backslash\{a,b\}} q_c =0$ or $1$, and 
$z[\{\rr_{c\to i};\mu_1]$ is defined as in Eq.~(\ref{RhoDefinition})
and discussion below. Notice that the above expression no longer depend upon 
$\mu_2$.
%
%
\subsection{$k$-SAT}
\label{ExplicitKSAT}

Once again, the first step consists in assigning the (model-dependent)
function $\rh^{\rm c}[\r^{(1)}\dots\r^{(k-1)}]$ entering in
Eq.~(\ref{TwoStep_2}). Unlike for $k$-XORSAT, this function depends 
upon the quenched variables $J_{a\to i}$. We have
\begin{eqnarray}
\rh^{\rm c}_{a\to i}(J_{a\to i}) & = &
\prod_{j\in\da\backslash i}\r_{j\to a}(-J_{a\to j}) \, ,
\label{KSAT_Cavity1}\\ 
\rh^{\rm c}_{a\to i}(0) & = & 1-
\prod_{j\in\da\backslash i}\r_{j\to a}(-J_{a\to j}) \, ,
\label{KSAT_Cavity2}\\
\rh^{\rm c}_{a\to i}(-J_{a\to i}) & = & 0\, ,
\label{KSAT_Cavity3}
\end{eqnarray}
where, for greater clarity, we used the notations $\r(q)$ and $\rh(q)$,
instead of $\r_q$ and $\rh_q$ for indicating the arguments of the 
distributions $\r$ and $\rh$. 
Obviously, the 1RSB solution (\ref{1RSBsolution}) has the property
that $\rr^{(q)}_{a\to i}=0$ for $q=-J_{a\to i}$. The distributions
$\rr_{a\to i}$ can therefore be parametrized by giving a single real number:
for instance $\rr^{(0)}_{a\to i}$.

Let us now consider the fluctuation parameters 
$\e^{(q)\sigma}_{i\to a}$ and $\eh^{(q)\sigma}_{a\to i}$. The fist important 
observation consists in noticing that, because of Eqs. 
(\ref{KSAT_Cavity1})--(\ref{KSAT_Cavity3}), 
$\eh^{(q)\sigma}_{a\to i}=0$ if $\sigma = -J_{a\to i}$ or $q=-J_{a\to i}$. 
Using the form of the transformation $T^{(a)}_{b\to i}$, this implies 
$\e^{(+)-}_{i\to a}=\e^{(-)+}_{i\to a}=0$. We are therefore left
with 4 parameters $\e_{i\to a}$'s and 2 $\eh_{a\to i}$'s different from zero.

In order to write explicit forms for the transformations 
(\ref{Linear1}) and (\ref{Linear2}), it is convenient to parametrize
the remaining $\e_{i\to a}$'s as follows:
\begin{eqnarray}
\e^{(+)+}_{i\to a} = (\ex^{\mu_1-2\mu_2}/r^{(+)}_{i\to a})\, 
\de^{(+)+}_{i\to a}\, ,&&\;\;\;
\e^{(0)+}_{i\to a} = (1/r^{(0)}_{i\to a})\, 
\de^{(0)+}_{i\to a}\, ,\\
\e^{(-)-}_{i\to a} = (\ex^{\mu_1-2\mu_2}/r^{(-)}_{i\to a})\, 
\de^{(-)-}_{i\to a}\, ,&&\;\;\;
\e^{(0)-}_{i\to a} = (1/r^{(0)}_{i\to a})\, 
\de^{(0)-}_{i\to a}\, .
\end{eqnarray}
Analogously we parametrize the non-zero $\eh^{(q)\sigma}_{a\to i}$'s 
in terms of $\deh^{(q)\sigma}_{a\to i}$'s.
We are now in position to write the explicit form of the transformations
(\ref{Linear1}), (\ref{Linear2}) in a compact form. In this case 
it is more convenient not to use the matrix notation. We get:
\begin{eqnarray}
\de^{(+)+}_{i\to a} & = & \sum_{b\in \Gamma_+} \,t^{(a)}_{b\to i}(0)\; 
\deh^{(+)+}_{b\to i}+ \sum _{b\in \Gamma_-}\, t^{(a)}_{b\to i}(+1)\;
\deh^{(0)-}_{b\to i}\, ,\label{LinearKSAT1}\\
\de^{(0)+}_{i\to a} & = & \sum_{b\in \Gamma_+} \,t^{(a)}_{b\to i}(0)\; 
\deh^{(0)+}_{b\to i}+ \sum _{b\in \Gamma_-}\, t^{(a)}_{b\to i}(+1)\;
\deh^{(-)-}_{b\to i}\, ,\\
\de^{(0)-}_{i\to a} & = & \sum_{b\in \Gamma_+} \,t^{(a)}_{b\to i}(-1)\; 
\deh^{(+)+}_{b\to i}+ \sum _{b\in \Gamma_-}\, t^{(a)}_{b\to i}(0)\;
\deh^{(0)-}_{b\to i}\, ,\\
\de^{(-)-}_{i\to a} & = & \sum_{b\in \Gamma_+} \,t^{(a)}_{b\to i}(-1)\; 
\deh^{(+)+}_{b\to i}+ \sum _{b\in \Gamma_-}\, t^{(a)}_{b\to i}(0)\;
\deh^{(-)-}_{b\to i}\, ,\label{LinearKSAT4}
\end{eqnarray}
where we used the shorthand $\Gamma_{\sigma} =$ 
$\{b\in\di\backslash a\, :\, J_{b\to i} = \sigma\}$. As for 
Eq.~(\ref{Linear2}), we get
\begin{eqnarray}
\deh^{(J_i)J_i}_{a\to i} & = & \sum_{j\in\da\backslash i}
\prod_{l\in\da\backslash\{ i,j\}}(1-r^{(0)}_{l\to a})\, 
\de_{j\to a}^{(-J_j)-J_j}\, ,\\
\deh^{(0)J_i}_{a\to i} & = & \sum_{j\in\da\backslash i}
\prod_{l\in\da\backslash\{ i,j\}}(1-r^{(0)}_{l\to a})\, 
\de_{j\to a}^{(0)-J_j}\, ,\label{LinearKSAT6}
\end{eqnarray}
where we used the shorthands $J_i, J_j$ for (respectively) $J_{a\to i}$
and $J_{a\to j}$. The last thing which remains to be specified are
the coefficients $t^{(a)}_{b\to i}(q)$ entering in 
Eqs. (\ref{LinearKSAT1})--(\ref{LinearKSAT4}):
\begin{eqnarray}
t^{(a)}_{b\to i}(q) & = &\frac{1}{z[\{\rr_{c\to i};\mu_1]}
\sum_{\stackrel{\{q_c\}}{\scriptscriptstyle \sum q_c = q}} 
\prod_{c\in\di\backslash\{a,b\}}(\rr^{(q_c)}_{c\to i} \ex^{-\mu_1|q_c|})\, .
\end{eqnarray}

As for $k$-XORSAT, the recursions (\ref{LinearKSAT1})--(\ref{LinearKSAT6})
can be further simplified by restricting them to a particular linear subspace.
Notice in fact that the subspace $\de^{(+)+}_{i\to a} = 
\de^{(0)+}_{i\to a}$, $\de^{(-)-}_{i\to a} = \de^{(0)-}_{i\to a}$,
$\deh^{(J_{a\to i})J_{a\to i}}_{a\to i} = 
\deh^{(0)J_{a\to i}}_{a\to i}$, is preserved by the above transformation.
A numerical calculation confirms that it contains the largest eigenvalue.
A further simplification occurs as $\mu_1\to\infty$, since in 
this limit $t^{(a)}_{b\to i}(\pm1)\to 0$
%
%
\section{$k$-XORSAT: expansion near the dynamic transition}
\label{ExpansionXORSAT}

In this Appendix we focus on XORSAT and expand both $e_{\rm d}(\alpha)$ 
and  $e_{\rm G}(\alpha)$ for $\alpha\to\alpha_{\rm d}$. This provides
us with an analytic characterization of the dynamic transition.

Throughout this Section, we shall work within the random $k$-XORSAT
{\it ensemble} defined in Sec.~\ref{NumericalSection}.
%
%
\subsection{Dynamic energy: $e_{\rm d}(\alpha)$}

Before dwelling upon the calculation, let us recall some well known
relations valid within a 1RSB approximation. The complexity can be
obtained as the Legendre transform of the replicated free energy
\cite{MonassonMarginal}:
\begin{eqnarray}
\Sigma(e)  =  \mu e - \mu\phi(\mu)\, ,\;\;\;\;\;
e =  \frac{\partial\phantom{\mu}}{\partial\mu}[\mu\phi(\mu)]\, .
\label{LegendreFormula}
\end{eqnarray}
Since $e_{\rm d}(\alpha)\downarrow 0$ as $\alpha\downarrow\alpha_{\rm d}$,
we are interested in the zero-energy (large $\mu$) 
limit of the above expressions. In this limit the free energy admits an 
expansion of the form:
\begin{eqnarray}
\mu\phi(\mu) = \phi_0 + \phi_1\, \ex^{-2\mu} +  \phi_2\, \ex^{-4\mu} +
O(\ex^{-6\mu})\, .\label{ExpansionForm}
\end{eqnarray}
which implies
\begin{eqnarray}
\Sigma(e)  = -\phi_0 -\frac{1}{2}\, e\, \log\left(\frac{e}{-2\phi_1}\right)+
\frac{1}{2}\, e-\frac{\phi_2}{4\phi_1^2}e^2+O(e^3)\, .
\end{eqnarray}
The dynamical energy (and the corresponding replica-symmetry breaking 
parameter) is defined as the location of the maximum of the complexity.
It is therefore easy to obtain
\begin{eqnarray}
e_{\rm d} = \frac{\phi_1^2}{4\phi_2}+O(\phi_1^4/\phi_2^2)\, ,\;\;\;\;\;\;\;\;
\mu_{\rm d} = -\frac{1}{2}\log\left(\frac{-\phi_1}{8\phi_2}\right)+
O(\phi_1^2/\phi_2)\, .\label{ExpansionEdGeneral}
\end{eqnarray}

Let us now turn to the case of random $k$-XORSAT. 
The 1RSB variational free energy reads
\begin{eqnarray}
\mu\phi(\mu) & = & k\alpha \int\!\! dP(r)\!\int\!\! d\Ph(\rr)\,\, \log\left[
1+\frac{1}{2}(\ex^{-2\mu}-1)(1-r_0)(1-\rr_0)\right]-\nonumber\\
&&-\alpha\int\!\!\prod_{i=1}^k dP(r^{(i)})\,\, \log\left[
1+\frac{1}{2}(\ex^{-2\mu}-1)\prod_{i=1}^k(1-r^{(i)}_0)\right]-\nonumber\\
&&-\sum_{l=0}^{\infty} p_l\int\!\!\prod_{i=1}^l d\Ph(\rr^{(i)})\,\, \log
\left[\sum_{q_1\dots q_l}\prod_{i=1}^l\rr^{(i)}_{q_i}\, 
\ex^{-\mu\sum_{i}|q_i|+\mu|\sum_i q_i|}\right]\, ,\label{XorSatFreeEnergy}
\end{eqnarray}
where $p_l = \ex^{-k\alpha}(k\alpha)^l/l!$ is the connectivity distribution of 
the variable nodes in the factor graph. The order parameters 
$r$ and $\rr$ are symmetric distributions over $\{+,0,-\}$ (which can therefore
parametrized using a single real number) and represent the distribution of 
the cavity fields. The functions $P(r)$ and $\Ph(\rr)$ are their 
distributions with respect to the disorder.

As shown in Refs. \cite{CoccoEtAlXorSat} and \cite{MezardEtAlXorSat} it
is important to distinguish the ``core'' of the factor graph. Outside
the core the cavity fields are trivial: $r_q = \rr_q = \delta_{q,0}$.
We rewrite \cite{MezardEtAlXorSat}:
\begin{eqnarray}
P[r] & = & u\,F[r]+(1-u)\,\delta[r-\delta^{(0)}]\, ,\label{Core1}\\
\Ph[\rr] & = & \uh\,\Fh[\rr]+(1-\uh)\,\delta[\rr-\delta^{(0)}]\, ,\label{Core2}
\end{eqnarray}
where the parameters $u$ and $\uh$ satisfy the self-consistency equations
\begin{eqnarray}
\uh = u^{k-1}\, ,\;\;\;\; u = 1-\ex^{-k\alpha \uh}\, .
\end{eqnarray}
These equations have a non-trivial solution $u,\uh>0$ for 
$\alpha \ge\alpha_{\rm d}$.

Plugging the decomposition (\ref{Core1}), (\ref{Core2}) into
Eq.~(\ref{XorSatFreeEnergy}) we get:
\begin{eqnarray}
\mu\phi(\mu) & = & k\alpha u\uh
\int\!\! dF(r)\!\int\!\! d\Fh(\rr)\,\, \log\left[
1+\frac{1}{2}(\ex^{-2\mu}-1)(1-r_0)(1-\rr_0)\right]-\nonumber\\
&&-\alpha u^k\int\!\!\prod_{i=1}^k dF(r^{(i)})\,\, \log\left[
1+\frac{1}{2}(\ex^{-2\mu}-1)\prod_{i=1}^k(1-r^{(i)}_0)\right]-\nonumber\\
&&-(1-\ex^{-k\alpha\uh})
\sum_{l=1}^{\infty} f_l\int\!\!\prod_{i=1}^l d\Fh(\rr^{(i)})\,\, \log
\left[\sum_{q_1\dots q_l}\prod_{i=1}^l\rr^{(i)}_{q_i}\, 
\ex^{-\mu\sum_{i}|q_i|+\mu|\sum_i q_i|}\right]\, ,\label{XorSatFreeEnergy_Core}
\end{eqnarray}
where $f_l = (\ex^{k\alpha\uh}-1)^{-1}(k\alpha\uh)^l/l!$ is the 
connectivity distribution inside the core. The 1RSB saddle-point equations 
can be obtained by differentiating the above expression with respect to 
the distributions $F(r)$ and $\Fh(\rr)$:
\begin{eqnarray}
F(r) & = & \sum_{l=1}^{\infty} f_l\int\!\!\prod_{i=1}^l d\Fh(\rr^{(i)})
\;\delta[r-\r^{\rm c}[\rr^{(1)}\dots\rr^{(l)}]]\, ,\label{SaddleCore1}\\
\Fh(\rr) & = & \int\!\!\prod_{i=1}^{k-1} dF(r^{(i)})\; 
\delta[\rr - \r^{\rm c}[r^{(1)}\dots r^{(k-1)}]]\, .\label{SaddleCore2}
\end{eqnarray}
The saddle point equations imply
that, as $\mu\to\infty$, $F(r)$ and $\Fh(\rr)$ are supported over 
$r_0, \rr_0 = O(\ex^{-2\mu})$. In particular we get
\begin{eqnarray}
\<r_0\>_F = \frac{f_2}{1-(k-1)f_1}\, \ex^{-2\mu}+O(\ex^{-4\mu})\, ,
\;\;\;\;\;\;\;
\<\rr_0\>_{\Fh} = \frac{(k-1)f_2}{1-(k-1)f_1}\,\ex^{-2\mu}+O(\ex^{-4\mu})\, .
\label{Average_r}
\end{eqnarray}

Using these results, one can expand Eq.~(\ref{XorSatFreeEnergy_Core})
for $\mu\to\infty$. We get the form (\ref{ExpansionForm}) with (the
first two coefficients were already derived in
\cite{MezardEtAlXorSat}):
\begin{eqnarray}
\phi_0(\alpha) & = &-[k\alpha u\uh-\alpha u^k-k\alpha\uh+(1-\ex^{-k\alpha\uh})
]\log 2\, ,\\
\phi_1(\alpha) & = & -\alpha u^k+\frac{1}{2}(k\alpha\uh)^2\,
\ex^{-k\alpha\uh}\, ,\\
\phi_2(\alpha) & = &
\frac{1}{2}\alpha u^k-\frac{1}{4}\,(k\alpha\uh)^2 \, \ex^{-k\alpha\uh}
\left[ 1- 2 k\alpha\uh+ \frac{1}{2}(k\alpha\uh)^2\right]+\nonumber\\
&&+\frac{1}{4}(k\alpha\uh)^5\frac{(k-1)(1-u)^2}{u[1-
k(k-1)\alpha(1-u)u^{k-2}]}\, .
\end{eqnarray} 
It is easy to check that both $\phi_0(\alpha)$ and $\phi_1(\alpha)$
have finite limits $\phi_{0,{\rm d}}$ and  $\phi_{1,{\rm d}}$ as 
$\alpha\downarrow\alpha_{\rm d}$. On the other hand, in the same limit,
$\phi_2(\alpha) = \phi_{2,{\rm d}} (\alpha-\alpha_{\rm d})^{-1/2}+
O(1)$, with
\begin{eqnarray}
\phi_{2,{\rm d}} = \frac{1}{4}(k\alpha_{\rm d}\uh_{\rm d})^4
\frac{(k-1)\alpha_{\rm d}(1-u_{\rm d})}{\sqrt{2(k-1)\alpha_{\rm d}
[k(k-1)\alpha_{\rm d}\uh_{\rm d}-(k-2)]}}\, ,
\end{eqnarray} 
where we defined $u_{\rm d}=\lim_{\alpha\to\alpha_{\rm d}}u$,
$\uh_{\rm d}=\lim_{\alpha\to\alpha_{\rm d}}\uh$.
Using these expressions in Eq.~(\ref{ExpansionEdGeneral}), 
we recover the first of the two results in (\ref{ExpansionMainText}) with 
$e^{(0)}_{\rm d} = \phi_{1,{\rm d}}^2/(4 \phi_{2,{\rm d}})$. Moreover
\begin{eqnarray}
\mu_{\rm d}(\alpha) = -\frac{1}{4}\log(\alpha-\alpha_{\rm d})
-\frac{1}{2}\log\left(\frac{-\phi_{1,{\rm d}}}
{8\phi_{2,{\rm d}}}\right)+ O((\alpha-\alpha_{\rm d})^{1/2})\, .
\end{eqnarray} 

One can define one more instability threshold $\mu_{1\to 1}(\alpha)$
such that the 1RSB solution becomes unstable within the 1RSB space
itself for $\mu<\mu_{1\to 1}(\alpha)$.  For $\mu<\mu_{1\to
1}(\alpha)$, the 1RSB saddle point equations cannot be any longer
solved by a population dynamics algorithm.  One has, obviously,
$\mu_{1\to 1}(\alpha)<\mu_{\rm G}(\alpha)$. More surprisingly, for
$\alpha \gtrsim \alpha_{\rm d}$, $\mu_{\rm d}(\alpha) < \mu_{1\to
1}(\alpha)$. This implies that the first expression in
Eq.~(\ref{ExpansionMainText}) cannot be directly tested against the
results of a population dynamics calculation.
%
%
%
\subsection{Stability threshold: $e_{\rm G}(\alpha)$}

In order to compute the stability threshold, it is helpful to restrict
ourselves to the core as in the previous Section: the fluctuation
parameters $\e_{i\to a}$ and $\eh_{a\to i}$ vanish outside.  Moreover,
we can work with the ``reduced'' recursion (\ref{ReducedRecursion}).
It is natural to consider the behavior of the joint probability
distributions $F(r,\de)$, $\Fh(\rr,\deh)$ under the recursions
(\ref{1RSBsolution}) and (\ref{ReducedRecursion}).  In analogy with
Eqs. (\ref{SaddleCore1}), (\ref{SaddleCore2}), we obtain
\begin{eqnarray}
F(r,\de) & = & \sum_{l=1}^{\infty} f_l\int\!\!\prod_{i=1}^l 
d\Fh(\rr^{(i)},\deh^{(i)})
\;\delta[r-\r^{\rm c}[\{\rr^{(j)}\}]]\;
\delta [\de-\sum_{i=1}^l t^{(i)}[\{\rr^{(j)}\}]\deh^{(i)} ]\, ,
\label{Fdelta1}\\
\Fh(\rr,\deh) & = & \int\!\!\prod_{i=1}^{k-1} dF(r^{(i)},\de^{(i)})\; 
\delta[\rr - \r^{\rm c}[\{r^{(j)}\}]]\;
\delta [\deh-\sum_{i=1}^{k-1} \th^{(i)}[\{r^{(j)}\}]\de^{(i)} ]\, ,
\label{Fdelta2}
\end{eqnarray}
where the coefficients $t^{(i)}[\dots]$ and $\th^{(i)}[\dots]$ are
easily computed using Eqs. (\ref{Coefficients1}) and
(\ref{Coefficients2}).  Alternatively these equations could have been
derived by ``projecting'' Eqs. (\ref{TwoStep_1}) and
(\ref{TwoStep_2}).

Notice that the marginal distributions $F(r)$ and $\Fh(\rr)$ satisfy
the Eqs. (\ref{SaddleCore1}), (\ref{SaddleCore2}). Therefore, if
$\mu\to\infty$, $F(r,\de)$, $\Fh(\rr,\deh)$ are supported on $r_0,
\rr_0 = O(\ex^{-2\mu})$. As for $\de$, $\deh$ two cases are possible:
either their support shrinks to $0$ (and therefore the 1RSB solution
is stable) or it remains distinct from $0$. This can be checked by
looking at the average value of $\de$, $\deh$ with respect to the
above distributions.  Using Eqs. (\ref{Fdelta1}), (\ref{Fdelta2}), we
get
\begin{eqnarray}
\<\de\>_F  =  \sum_{l=1}^{\infty} \sum_{i=1}^l \, f_l\,
\<t^{(i)}[\{\rr^{(j)}\}]\cdot \deh^{(i)}\>_{\Fh}\, ,\;\;\;\;\;\;
\<\deh\>_{\Fh}  =  \sum_{i=1}^{k-1} \<\th^{(i)}[\{r^{(j)}\}]\cdot 
\de^{(i)}\>_F\, .
\end{eqnarray}

We are interested in the regime $\alpha\downarrow \alpha_{\rm d}$,
$\mu_1\uparrow\infty$, with $\ex^{-\mu_1}\sim (\alpha-\alpha_{\rm d})$.
This can be checked to be the correct scaling at the end of the computation.
If we expand the above equations in this limit, we get
\begin{eqnarray}
\<\de\>_F  & = &  [f_1+2f_2(\ex^{-\mu_1}+2 \<\rr_0\>_{\Fh})]\, 
\<\deh\>_{\Fh}+O(\ex^{-2\mu_1})\, ,\\
\<\deh\>_{\Fh}  & = & (k-1) [1-(k-2) \<r_0\>_{F}]\,
\<\de^{(i)}\>_F+O(\ex^{-2\mu_1})\, .
\end{eqnarray}
which imply the marginality condition
\begin{eqnarray}
(k-1)[1-(k-2) \<r_0\>_{F}] [f_1+2f_2(\ex^{-\mu_1}+2 \<\rr_0\>_{\Fh})]
= 1 +O(\ex^{-2\mu_1})\, .
\end{eqnarray}
Notice that $\<r_0\>_{F}$ and $\<\rr_0\>_{\Fh}$ are formally of order
$\ex^{-2\mu_1}$, cf. Eq.~(\ref{Average_r}) but since
$1-(k-1)f_1=O(\alpha-\alpha_{\rm d})$, they must be in fact considered
of order $\ex^{-\mu_1}$.  The relation can be inverted yielding
\begin{eqnarray}
\mu_{\rm G}(\alpha) = -\frac{1}{2}\log(\alpha-\alpha_{\rm d}) -\log A +
O((\alpha-\alpha_{\rm d})^{1/2})\, ,
\end{eqnarray}
where
\begin{eqnarray}
A = \frac{2(1-u_{\rm d})}{\zeta\alpha_{\rm d}u_{\rm d}}\sqrt{
2(k-1)\alpha_{\rm d}[k(k-1)\alpha_{\rm d}\uh_{\rm d}-(k-2)]}\, ,
\;\;\;\;\;\zeta = 1+\sqrt{5-\frac{2(k-2)}{k(k-1)\alpha_{\rm d}\uh_{\rm d}}}\, .
\nonumber\\
\end{eqnarray}
Plugging this result into the general relations
(\ref{LegendreFormula}), (\ref{ExpansionForm}), we get the second
result in Eq.~(\ref{ExpansionMainText}), with $e^{(0)}_{\rm G} =
-2A^2\phi_{1,d}$.

%
%

\end{document}